\documentclass[9pt,a4paper,twocolumn,twoside]{rho-class/rho}
\usepackage[english]{babel}
\usepackage{graphicx}
\usepackage{subcaption}

\title{Gendered Structural Roles in Indian Epics}

\author[a]{J.D.P. Evans, University of Hertfordshire, \href{mailto:j.evans8@herts.ac.uk}{j.evans8@herts.ac.uk}}

\journalname{}
\journal{}
\theday{\today}

\begin{abstract}
    The \textit{Mah\={a}bh\={a}rata} and the \textit{R\={a}m\={a}ya\d{n}a} constitute the two major Sanskrit epics of ancient India. Previous network analysis revealed substantial structural similarities between their character networks, alongside differences consistent with their contrasting forms of narrative organisation. Building upon these findings, we investigate the representation and structural positions of male and female characters within the two texts. In particular, we examine whether the genders differ systematically in their connectivity, prominence, community participation, and contribution to network cohesion, while accounting for the considerable imbalance between the numbers of male and female characters. We employ methods designed to identify characters who connect otherwise distinct regions of the networks and to assess the structural consequences of character removal. Through these analyses, we ask whether gendered patterns are consistent across the two epics or instead reflect the different narrative structures previously identified. In so doing, we find that male characters are considerably more prevalent among highly central and structurally consequential nodes in both epics, and targeted removal of central male characters produces substantially greater network disruption. However, once baseline structural position is taken into account, we find little residual difference in deletion effects between male and female characters. Furthermore, under selected Girvan-Newman partitions, we find that female characters in the \textit{Mah\={a}bh\={a}rata} exhibit greater cross-community participation than female characters in the \textit{R\={a}m\={a}ya\d{n}a}.
\end{abstract}


\begin{document}

    \maketitle
    \thispagestyle{firststyle}


\section{Introduction}
\label{sec:Introduction}

The use of complex networks has in recent years provided quantitative insights into existing problems in comparative mythology, such as historicity, genre, narrative structure, cultural transmission, character prominence, and gender representation, cf. \cite{Evans2026b}, \cite{JanickyjMacCarronKenna2024}, \cite{MacCarronKenna2013}, \cite{MacCarronKenna2012}, \cite{YoseMacCarron2018}, \cite{YoseMacCarron2016}. Much of the existing work has focussed on European mythology, such as those of Ireland \cite{MacCarronKenna2012}, \cite{YoseMacCarron2018}, Iceland \cite{MacCarronKenna2013}, Scotland \cite{YoseMacCarron2016}, Greece \cite{MacCarronKenna2012}. However, in recent years other mythologies have been considered, \cite{Evans2026b}, \cite{GultepeMathangi2023}. It remains unclear whether observed structural differences primarily reflect culture, genre, narrative scale, textual transmission, or differences in network construction. Nevertheless, the increasing range of traditions being studied indicates both growing interest in the field and the need for comparative analyses.

In \cite{Evans2026b} and \cite{GultepeMathangi2023}, network analysis was used to examine Indian mythological texts, albeit from different methodological viewpoints (see Section \ref{sec:network_construction}). Janickyj et al. included the \textit{Mah\={a}bh\={a}rata} in a cross-cultural study of gender differences in mythological character networks, \cite{JanickyjMacCarronKenna2024}. In \cite{Evans2026b}, we compared the \textit{Mah\={a}bh\={a}rata} and the \textit{R\={a}m\={a}ya\d{n}a}, also comparing both texts with the Greek epic, the \textit{Iliad}. Our goal was two-fold. First, we investigated whether their network structures supported the traditional distinction between the \textit{Mah\={a}bh\={a}rata} as comparatively historical and the \textit{R\={a}m\={a}ya\d{n}a} as comparatively literary; secondly, we asked if there was structural evidence that the Indian texts were unsuitable for the predominantly Western label of `epic'. In both cases, we found insufficient structural evidence to support either claim.

With such questions of genre and label aside, the present study aims to continue and extend the comparisons of the Indian texts found in \cite{Evans2026b}, this time focusing on gender differences in the structural positions occupied by characters within the two narrative networks. Gender was also the primary focus of \cite{JanickyjMacCarronKenna2024}, in which Janickyj et al. utilised a variety of techniques to compare several mythological texts. They examined the extent to which female characters occupied more or less prominent structural positions across narrative traditions. For this study, our goals are quite different, not least because both texts emerge from related, although distinct, Sanskrit literary and religious traditions. As such, their comparison does not provide the same kind of cross-cultural contrast examined by Janickyj et al. Moreover, the long and complex transmission histories of both works mean that the present datasets cannot support a controlled comparison between clearly defined historical periods. Instead, we use gender to further our comparisons between the narrative networks of the \textit{Mah\={a}bh\={a}rata} and the \textit{R\={a}m\={a}ya\d{n}a}. While \cite{Evans2026b} did not find sufficient evidence to assign the two texts distinct genres, we did find several key structural differences between the texts, summarised in Section \ref{sec:epic_sanskrit_texts}. In particular, we found evidence consistent with narrative differences between the two texts, thereby raising the tantalising prospect that broad narrative themes may possibly be classified using the underlying network. The present study therefore investigates whether gendered differences in character representation and network position are consistent with the broader structural and narrative contrasts identified in our previous work. In addition to comparisons based on descriptive statistics and standard centrality measures, we introduce methods designed to examine the structural functions performed by female characters, including their potential roles as bridges between otherwise separated regions of the networks. One key question we shall be attempting to answer is the following: within each text, do male and female characters differ systematically in their connectivity, centrality, community participation, and contribution to global network cohesion, after accounting for their unequal representation? We further ask, are the resulting gendered patterns similar across the two texts, or do they reflect the different forms of narrative organisation identified in \cite{Evans2026b}?

In Section \ref{sec:network_construction}, we shall explain the construction of these networks. In Section \ref{sec:epic_sanskrit_texts}, we shall present a brief overview of the texts to contextualise what follows, alongside a summary of what was discovered in \cite{Evans2026b}. In Section \ref{sec:gender_roles}, we shall present the analysis itself. We conclude in Section \ref{sec:conclusion}. As with \cite{Evans2026b}, we shall include a brief summary of Sanskrit pronunciation in Section \ref{app:pronouncing_sanskrit}.

\section{Network Construction}
\label{sec:network_construction}

We construct the networks using the same approach as that of \cite{Evans2026b}. We will therefore repeat the description from our earlier study. Construction of the respective networks began with careful reading of the texts. Each character is represented as a node, alongside other information including the character's gender. Edges are then constructed when two characters interact physically in some way, or are related either by blood or marriage. Physical interactions can include speaking to each other directly or being part of a group in which it is clear that the two characters know each other. This allows for situations in which a group of characters are present but only a strict subset speak in the text, and avoids situations such as two characters both being present at an event (such as a celebration) but in which they do not actually meet or interact. Additionally, we differentiate between hostile and friendly interactions\footnote{We shall use interchangeably the terms positive/negative and friendly/hostile, respectively.}. We define a hostile interaction to be any interaction in which there is some form of violence. Primarily, this is physical but could also include other forms of violence such as a threat. A friendly interaction is then defined to be any other type of interaction. Repeated interactions are counted with each edge storing separate friendly and hostile interaction counts, denoted $w_{pos}$ and $w_{neg}$, respectively. Note that all interactions, friendly or hostile, are symmetric and so our networks will necessarily be undirected.

\section{Epic Sanskrit Texts}
\label{sec:epic_sanskrit_texts}

The \textit{Mah\={a}bh\={a}rata} is an Indian epic, written sometime between 400 BCE-400 CE\footnote{Note that ancient Indian chronology is notoriously difficult to establish and the \textit{Mah\={a}bh\={a}rata} is no exception.}. At its most basic, it is a story about a conflict between two sets of cousins. It includes the events which led to war, the eighteen-day war itself, and the aftermath of this war. Traditionally, it is said to contain 100,000 verses with most verses consisting of two lines each. A critical shortened edition is the Poona edition (published between 1933 and 1966), which retains 67,314 verses and is still considered to be authentic. For this work, we use the heavily abridged translation by J.D. Smith (see \cite{SmithMaha}) which is based upon the Poona edition. Given the heavily condensed nature of Smith's translation, all analysis which follows should keep this in mind. To give a sense of scale, in S. S\"{o}rensen's \textit{An Index to the Names in the Mahabharata} \cite{Sorensen1978}, the list runs to 807 two-column pages. In contrast, our network has just 536 characters. A justification for the use of this version can be found in \cite{Evans2026b}.

The other text of interest is the parallel tale of R\={a}ma. We shall use an early form of the \textit{R\={a}m\={a}ya\d{n}a} \cite{BrockingtonRama} translated by J. Brockington and M. Brockington called \textit{R\={a}m\={a} the Steadfast}. Likely to have originated as an oral composition (cf. \cite{BrockingtonRama}), the date of its composition is still a matter of debate. J. and M. Brockington suggest a date of roughly the Fifth Century BCE\footnote{This follows from linguistic evidence based upon texts which can be roughly dated. A date of First Century CE is reasonable for the written date.}, although many additions and changes took place over the following centuries, as was the case for the \textit{Mah\={a}bh\={a}rata}. Our translation comes from the earliest levels of the story and represents an early stage of its development. In the Introduction to \textit{R\={a}m\={a} the Steadfast}, the translators compare it to a Western heroic romance. In it, we meet the hero R\={a}ma, now deified following centuries of Indian tradition. Forced into exile by his father at the behest of his father's wife (not R\={a}ma's mother), he is joined by his brother Lak\d{s}ma\d{n}a and wife S\={i}t\={a}. During their exile, S\={i}t\={a} is kidnapped by the evil R\={a}va\d{n}a and the remainder of the text follows the ensuing war to rescue her. In contrast to the epic scope of the \textit{Mah\={a}bh\={a}rata}, the struggle here is that for integrity and happiness for R\={a}ma. In contrast to Arjuna's famous grappling with what action is permissible for a hero to follow while maintaining integrity, R\={a}ma's struggle is intensely personal. He wishes to fulfil his duty as a son by accepting his father's banishment, and then he wishes to rescue his wife. To once more quote the aforementioned Introduction, ``Indian tradition too distinguishes between the two works, designating the \textit{R\={a}m\={a}ya\d{n}a} as the \textit{\={a}dik\={a}vya}, `the first poetic work' or perhaps `the first work of pure literature', but the \textit{Mah\={a}bh\={a}rata} commonly as \textit{itih\={a}sa}, `thus indeed it was', a term roughly equivalent to history." Again, a justification for the use of this version can be found in \cite{Evans2026b}.

We repeat a warning from the conclusion of \cite{Evans2026b}. Both translations chosen involve a heavy amount of abridgement. It is an open problem to what extent different translations affect structural properties. Once different versions and/or abridgements are entered into the mix, the questions of representativeness of all results we obtain only increase. As such, we see a future direction involving a thorough analysis of networks relating to various translations and, most importantly, fuller versions of the text. We should stress the monumental effort such an undertaking would represent, owing to the continued addition and modification of the texts upon copying. For example, while the oldest \textit{R\={a}m\={a}ya\d{n}a} manuscript dates from 1020 CE, copies were formed as late as the Seventeenth and Eighteenth Centuries and they bear evidence of R\={a}m\={a}'s increasing veneration. A tantalising possibility may therefore be to explore the structural differences between the two main recensions which J. and M. Brockington describe as roughly corresponding to a Southern group and a Northern group. While the Southern group is traditionally regarded as remaining closer to the \textit{R\={a}m\={a}ya\d{n}a}'s origins, the degree of cross-fertilization within and between recensions and sub-recensions is far from insignificant.

To conclude this section, we briefly summarise the main results obtained in \cite{Evans2026b}. As ever, the reader is directed to the original study for more detail. Overall, we found that the two Indian texts exhibited more structural similarities than differences. Both networks displayed broadly small-world organisation, high levels of clustering, meaningful community structure, and robustness to the random removal of characters, while being considerably more vulnerable to targeted removal, particularly when characters were removed according to betweenness centrality. These common features suggested that both texts possess many of the structural characteristics associated with real-world social networks, although both are disassortative, a property not typically associated with such networks. We also saw low numbers of triangles with odd numbers of hostile edges, although the \textit{R\={a}m\={a}ya\d{n}a} does have substantially fewer such triads.

Nevertheless, several important differences emerged in the organisation of the two Indian networks. The \textit{Mah\={a}bh\={a}rata} exhibited a more intricate pattern of interaction among its highly connected characters, with interconnected elite groups acting as bridges between partially distinct narrative communities. This structure was consistent with the overlapping loyalties and affiliations produced by its civil war narrative. By contrast, the \textit{R\={a}m\={a}ya\d{n}a} was more locally cohesive, with influential characters more frequently embedded within tightly connected neighbourhoods and with a clearer structural division between the groups surrounding R\={a}m\={a} and R\={a}va\d{n}a. We argued that these differences were more plausibly interpreted as consequences of differing narrative organisation than as evidence for a strict historical/poetic distinction between the texts.

\section{Gender Roles in the Epics}
\label{sec:gender_roles}

In this section, we now turn to the role of gender in the Indian texts. As with many mythological texts, the \textit{Mah\={a}bh\={a}rata} and the \textit{R\={a}m\={a}ya\d{n}a} come from patriarchal cultures. Aside from the primary role of men in the narratives, an example of how this manifests in the texts themselves is through naming conventions. It is common for characters to identify themselves by the father's name by modifying the first and final syllables, e.g. S\={i}t\={a} also being called J\={a}nak\={i} (daughter of Janaka), or Draupad\={i}, daughter of Drupada, cf. Section \ref{app:pronouncing_sanskrit}. With such a view, and with larger than life male characters, it is natural to explore how femininity is constructed within these narratives and whether they exhibit their own agency.

\noindent\textbf{Question:} Are the female characters passive figures within the texts, or do they actively shape the plot?

Network structure alone cannot determine literary agency, motivation, or autonomy. However, we can address the narrower question of whether female characters occupy structurally prominent or consequential positions within interaction networks. The above question can therefore be seen to encompass three distinct aspects:
\begin{enumerate}
    \item \textbf{Representation:} How many male and female characters are present?
    \item \textbf{Structural position:} Do their centralities, connectivity, or community relationships differ?
    \item \textbf{Structural function:} Do they contribute differently to cohesion, bridging, or fragmentation?
\end{enumerate}

\begin{table}[H]
\caption{Percentage of men and women.}
\label{tab:gender_perc}

\centering
\begin{tabular}{l|cc}
\toprule
 & \textit{Mah\={a}bh\={a}rata} & \textit{R\={a}m\={a}ya\d{n}a} \\
\midrule
Male nodes & 443 & 208 \\
Female nodes & 89 & 39 \\
Percentage of men & 82.65\% & 84.21\% \\
Percentage of women & 16.60\% & 15.79\%  \\
\bottomrule
\end{tabular}

\tabletext{Note: In the \textit{Mah\={a}bh\={a}rata}, four collective nodes are not assigned a male/female gender label.}
\end{table}

\medskip

\noindent\textbf{Initial Statistics}
\medskip

\tabref{tab:gender_perc} provides an initial summary. In the first two sections of \tabref{tab:well_connected_women}, we gain further insight into how well-connected women are by looking at the percentage of men and women whose degrees are larger than the mean. We choose the mean degree since, as we saw in \cite{Evans2026b}, the \textit{R\={a}m\={a}ya\d{n}a} is well-fitted by an exponential degree distribution, whereas the \textit{Mah\={a}bh\={a}rata} is well-fitted by a Weibull distribution. Both distributions have a parameter related to the mean degree and  are relatively light-tailed compared with a power law, meaning the mean is reasonably representative. Consequently, we view the comparison as broadly suitable. For betweenness centrality, we again use the network-wide mean as a comparison. However, betweenness centrality distributions are generally highly skewed, which means the mean is less suitable. Nevertheless, the mean provides a simple and consistent network-wide benchmark for comparative purposes across texts and genders.

We see that both Indian texts have broadly similar gender compositions, with male characters substantially outnumbering female characters. A larger proportion of male than female characters lies above the network-wide mean for both degree and betweenness centrality, indicating men are more likely to occupy structurally central positions in both networks. However, there is some nuance here. In terms of degree, the \textit{R\={a}m\={a}ya\d{n}a} has a larger absolute percentage-point gap (37.5\% men above the average compared to 17.95\% women, against 27.09\% compared to 12.36\%), while this pattern is reversed for betweenness centrality with \textit{Mah\={a}bh\={a}rata} showing more extreme differences (18.74\% men above the average compared to 8.99\% women, against 17.79\% compared to 12.82\%). This suggests that in the \textit{R\={a}m\={a}ya\d{n}a}, male characters are more disproportionately represented among above-average-degree nodes, whereas in the \textit{Mah\={a}bh\={a}rata} male characters appear more disproportionately represented among characters with above-average betweenness.

\begin{table}[H]
\caption{Gender specific properties. The percentage of female (resp. male) nodes who have a degree greater than the network's average is denoted by $\%N_fk>\langle k\rangle$ (resp. $\%N_mk>\langle k\rangle$). The percentage of female (resp. male) nodes who have a betweenness centrality greater than the network's average is denoted by $\%N_fBC>\bar{BC}$ (resp. $\%N_mBC>\bar{BC}$). The average degree (resp. betweenness centrality) is denoted $\langle k\rangle$ (resp. $\bar{BC}$). We give absolute counts in parentheses.}
\label{tab:well_connected_women}

\centering
\begin{tabular}{c|cc}
\toprule

 & \textit{Mah\={a}bh\={a}rata} & \textit{R\={a}m\={a}ya\d{n}a} \\
\midrule

$\langle k\rangle$ & 9.9440 & 10.1700 \\
$\%N_fk>\langle k\rangle$ & 12.36 (11/89) & 17.95 (7/39) \\
$\%N_mk>\langle k\rangle$ & 27.09 (120/443) & 37.50 (78/208) \\

\midrule

$\bar{BC}$ & 0.0045 & 0.0045 \\
$\%N_fBC>\bar{BC}$ & 8.99 (8/89) & 12.82 (5/39) \\
$\%N_mBC>\bar{BC}$ & 18.74 (83/443) & 17.79 (37/208) \\

\midrule

F in top 10\% w.r.t. Degree (\%) & 9.26 (5/54) & 4 (1/25) \\
M in top 10\% w.r.t. Degree (\%) & 90.74 (49/54) & 96 (24/25) \\
F in top 10\% w.r.t. BC (\%) & 5.56 (3/54) & 12 (3/25)\\
M in top 10\% w.r.t. BC (\%) & 94.44 (51/54) & 88 (22/25) \\

\midrule 

Pct. F in top 10\% (Degree) & 5.62 (5/89) & 2.56 (1/39)\\
Pct. M in top 10\% (Degree) & 11.06 (49/443) & 11.54 (24/208) \\
Pct. F in top 10\% (BC) & 3.37 (3/89) & 7.69 (3/39) \\
Pct. M in top 10\% (BC) & 11.51 (51/443) & 10.58 (22/208) \\

\bottomrule
\end{tabular}
\end{table}
The last two blocks of \tabref{tab:well_connected_women} focus on how many women are in the top 10\% with respect to degree and betweenness centrality. This gives us an indication of gender representation among elite nodes. Focusing first on the proportion of men and women within the top 10\%, we see men are overwhelmingly represented within this metric. For the \textit{Mah\={a}bh\={a}rata}, 90.74\% of the top 10\% are men (with respect to degree), while 94.44\%  of the top 10\% are men (with respect to betweenness centrality). It is a similar story with the \textit{R\={a}m\={a}ya\d{n}a} with 96\% of the top 10\% being men (with respect to degree), while 88\% of the top 10\% are men (with respect to betweenness centrality). Note once again that in the \textit{R\={a}m\={a}ya\d{n}a} men are slightly less dominant with respect to betweenness centrality, compared to the \textit{Mah\={a}bh\={a}rata}.

However, the above could simply reflect the severe male/female imbalance in the two texts. For this reason, we next consider percentage of men and women in the respective top 10\%. This normalises by gender-group size and so provides a more informative comparison. Again, the same pattern of substantial gender asymmetry in structural prominence is clearly evident. The \textit{Mah\={a}bh\={a}rata} sees men dominate with respect to both measures, though their dominance is more extreme when considering betweenness centrality. For the \textit{R\={a}m\={a}ya\d{n}a}, this picture is reversed and men dominate more extremely when it comes to degree.

At this point, there is a relatively clear, if unsurprising, picture of male domination, albeit one in which the form this dominance takes is slightly different between the texts. The \textit{Mah\={a}bh\={a}rata} exhibits a stronger concentration of betweenness centrality among male characters. In contrast, although women remain numerically marginal in the \textit{R\={a}m\={a}ya\d{n}a}, they appear relatively more visible in structurally intermediary positions, as reflected in the betweenness centrality. This broadly aligns with the arguments of \cite{Evans2026b} that the \textit{R\={a}m\={a}ya\d{n}a} exhibits stronger local cohesion overall, while the \textit{Mah\={a}bh\={a}rata} more frequently displays structurally intermediary characters that potentially occupy brokerage-like structural positions. Within this framework, female characters in the \textit{R\={a}m\={a}ya\d{n}a} may occupy comparatively more integrative narrative roles despite limited overall representation. To explore the validity of this possibility, we therefore need to dig deeper. In particular, we want to understand how women interact throughout the network.
\medskip

\noindent\textbf{Gender-Based Targeted Node Removal}

\begin{table}[H]
    \centering
    \caption{Edges with female characters. We use $E_1$ (resp. $E_2$) to denote the number of edges with at least one (resp. two) female character(s). Thus, $E_2\subseteq E_1$,
    and $E_1-E_2$ is the number of mixed-gender edges.}
    \begin{tabular}{c|cccc}
    \toprule
        & $E_1$ & $E_1$ (\%) & $E_2$ & $E_2$ (\%) \\
    \midrule
        \textit{Mah\={a}bh\={a}rata}  &  501 & 18.80 & 44 & 1.65\\
         \textit{R\={a}m\={a}ya\d{n}a} & 191 & 15.21 & 57 & 4.54 \\
        \bottomrule
    \end{tabular}
    \label{tab:female_edges}
\end{table}

\tabref{tab:female_edges} shows that in both texts, edges between female characters form a very small proportion of the total edge set. Although mixed-gender edges are considerably more numerous than female/female edges, the edge sets of both networks consist overwhelmingly of male/male interactions. A further distinction emerges when we separate mixed-gender edges from female/female edges. In the \textit{Mah\={a}bh\={a}rata}, the proportion of mixed-gender edges is $\frac{501-44}{2665}\times 100 \approx 17.15\%$, compared with $\frac{191-57}{1256}\times 100 \approx 10.67\%$ in the \textit{R\={a}m\={a}ya\d{n}a}. In contrast, female/female edges form a larger proportion of the \textit{R\={a}m\={a}ya\d{n}a} network, accounting for 4.54\% of all edges compared with 1.65\% in the \textit{Mah\={a}bh\={a}rata}. Equivalently, conditional on an edge involving at least one female character, the probability that both endpoints are female is $\frac{44}{501}\times 100 \approx 8.78\%$ for the \textit{Mah\={a}bh\={a}rata}, but $\frac{57}{191}\times 100 \approx 29.84\%$ for the \textit{R\={a}m\={a}ya\d{n}a}. Thus, although a slightly smaller proportion of edges in the \textit{R\={a}m\={a}ya\d{n}a} involve female characters, those that do are substantially more likely to connect two women.

Of course, what we wish to understand is how influential these edges involving women actually are upon the network as a whole. The above describes the prevalence of edges involving female characters, but not the structural importance of the characters incident to them. We therefore examine how the networks respond to the targeted removal of highly central male and female characters. We first compute betweenness centrality in the complete network, after which male and female characters are ranked separately. We then remove the top 5\% of female nodes (ranked within females according to betweenness centrality) and track the following:
\begin{itemize}
    \item Relative giant component size;
    \item Number of connected components;
    \item Average path length of the giant component;
    \item Edge loss.
\end{itemize}
We then repeat this for the top 5\% of male nodes (ranked within males according to betweenness centrality), and compare. Since female characters are numerically under-represented in both networks, comparing the most central nodes overall would largely reproduce the male dominance already discussed. By ranking nodes within gender and removing the top $5\%$ separately, we attempt to accommodate this. However, because the male and female node sets differ substantially in size, this should be interpreted as a proportional within-gender attack rather than an equal-count removal.

\begin{table*}[t]
\centering
\caption{Effects of removing the top $5\%$ most central female and male nodes, ranked within gender by betweenness centrality: $\Delta GC$ denotes the change in giant component size; $\Delta GC(pp)$ denotes the percentage point change in the proportion of nodes lying in the giant component; $\Delta E_{GC} (pp)$ denotes the percentage point change in the proportion of edges within the giant component; $\Delta l_{GC}$ denotes the change in average path length within the giant component; $\Delta C$ denotes the change in number of connected components within the network.}
\label{tab:gender_attack_summary}

\begin{tabular}{l|l|cccccccc}
\toprule
\textbf{Network} & \textbf{Group removed} &
$n_{\mathrm{removed}}$ &
$\Delta$GC &
$\Delta$GC (pp) &
Edge loss &
Edge loss (\%) &
$\Delta E_{\mathrm{GC}}$ (pp) &
$\Delta \ell_{\mathrm{GC}}$ &
$\Delta C$ \\
\midrule

\textit{Mah\={a}bh\={a}rata}
& Top females
& 5
& -10
& -0.97
& 137
& 5.14
& -0.18
& -0.04
& +2 \\

& Top males
& 23
& -138
& -22.56
& 1112
& 41.73
& -3.81
& +1.61
& +77 \\

\midrule

\textit{R\={a}m\={a}ya\d{n}a}
& Top females
& 2
& -4
& -0.97
& 49
& 3.90
& -0.04
& +0.02
& +2 \\

& Top males
& 11
& -55
& -19.49
& 447
& 35.59
& -1.97
& +0.77
& +32 \\

\bottomrule
\end{tabular}
\end{table*}

The results are summarised in \tabref{tab:gender_attack_summary}. As expected given the difference in number of nodes removed, removing the most central male characters produces dramatically greater disruption than removing the most central female characters. In particular, we notice once again the \textit{Mah\={a}bh\={a}rata}'s fragility to attack based on betweenness centrality. The effect is especially pronounced for average path length, where the increase is approximately twice that observed in the \textit{R\={a}m\={a}ya\d{n}a}. More specifically for gender, we see that in the \textit{Mah\={a}bh\={a}rata}, removing the top 5\% of female nodes reduces the giant component by less than one percentage point, whereas removing the top 5\% of male nodes reduces it by more than 22 percentage points, an enormous disparity. Similarly, in the \textit{R\={a}m\={a}ya\d{n}a}, top-female-node removal reduces the giant component by roughly one percentage point, while top-male removal reduces it by almost 20 percentage points. Regarding the changes in average path length, highly central male characters appear to play a stronger role in maintaining short paths across the network, as compared to female characters. Among the nodes remaining in the post-removal giant component, shortest-path distances increase substantially following male removal but change little following female removal. Of course, the effects seen above could simply be because we are removing a greater number of male nodes. To see if that is the case, \tabref{tab:gender_attack_equal_count} looks at an equal count removal. The results are broadly the same, albeit the effects slightly less extreme. Nevertheless, even under equal-count removal, the most central male characters exert larger structural effects than the most central female nodes.

\begin{table*}[t]
\centering
\caption{Effects of removing equal numbers of the most central female and male nodes, ranked within gender by betweenness centrality. Columns are denoted according to the same conventions as \tabref{tab:gender_attack_summary}.}
\label{tab:gender_attack_equal_count}

\begin{tabular}{l|l|cccccccc}
\toprule
\textbf{Network} & \textbf{Group removed} &
$n_{\mathrm{removed}}$ &
$\Delta$GC &
$\Delta$GC (pp) &
Edge loss &
Edge loss (\%) &
$\Delta E_{\mathrm{GC}}$ (pp) &
$\Delta \ell_{\mathrm{GC}}$ &
$\Delta C$ \\
\midrule

\textit{Mah\={a}bh\={a}rata}
& Top females
& 5
& -10
& -0.97
& 137
& 5.14
& -0.18
& -0.04
& +2 \\

& Top males
& 5
& -43
& -7.19
& 448
& 16.81
& -0.53
& +0.49
& +30 \\

\midrule

\textit{R\={a}m\={a}ya\d{n}a}
& Top females
& 2
& -4
& -0.97
& 49
& 3.90
& -0.04
& +0.02
& +2 \\

& Top males
& 2
& -15
& -5.45
& 168
& 13.38
& -0.41
& +0.27
& +10 \\

\bottomrule
\end{tabular}
\end{table*}
\bigskip 

{}
\bigskip

\noindent\textbf{Participation Coefficients}
\medskip

Thus far we have simply measured global connectivity and influence. Roughly speaking, we have seen that female nodes are not, on average, the main supports in the giant component. However, female structural importance may simply exist elsewhere. With this in mind, let us ask the following question: Even if female nodes are not the main global backbone, are they disproportionately important as connectors between communities? This \textit{community brokerage} is what we now attempt to understand. We proceed by first partitioning each network using the Girvan-Newman community detection algorithm. For a node $i$, we then denote by $k_{is}$ the number of edges of node $i$ to nodes in community $s$. The \textit{participation coefficient} of \cite{GuimeraAmaral2005} is then defined to be,

\[
P_i
=
1-\sum_s
\left(
\frac{k_{is}}{k_i}
\right)^2.
\]

\noindent This measures how evenly a node distributes its connections across different communities. It is close to one if all a node's edges are uniformly distributed among all the communities, and zero if all the edges are within its own community. For descriptive purposes, we use the following heuristic interpretation:
\begin{itemize}
    \item Near 0 means almost all neighbours lie in one community. This node is community-embedded and is not much of a broker;
    \item 0.2-0.6 is a small-to-moderate value and means the node connects to more than one community but still has a clear home base;
    \item 0.6-0.8 is a moderate-to-high value and means the node is fairly well spread across communities and is acting more like a connector/broker;
    \item Very high values (0.8+) mean neighbours are distributed quite evenly across several communities. Such nodes fall within the range associated with `kinless' roles in the classification of Guimer\'{a} and Amaral, although we do not apply their full role classification because we do not use within-community degree.
\end{itemize}

Our intention with the above is to distinguish locally cohesive characters from structurally intermediary characters whose relationships span multiple communities. This distinction is particularly relevant in the present context, for in \cite{Evans2026b} we suggested that the \textit{Mah\={a}bh\={a}rata} exhibits weaker local cohesion among highly connected nodes, stronger fragmentation under targeted attack using betweenness centrality, and have now seen a greater gender asymmetry among structurally intermediary characters. In contrast, the \textit{R\={a}m\={a}ya\d{n}a} exhibits stronger local clustering and more cohesive community structure. The participation coefficient therefore provides a natural way to test whether these broader structural differences are reflected in the extent to which important characters act within communities or across them. In particular, high participation coefficients are associated with nodes whose narrative relationships span multiple communities, while low participation coefficients indicate nodes embedded primarily within a single cohesive faction or social circle. As such, this measure provides a complementary perspective to degree and betweenness centrality by distinguishing locally prominent nodes from nodes whose relationships span multiple communities.

Our approach does not fully match that of Guimer\'{a} and Nunes Amaral in \cite{GuimeraAmaral2005}, however. They used within-community degree and participation coefficient to classify hub/non-hub nodes\footnote{Guimer\'{a} and Nunes Amaral assigned non-hub nodes to one of four roles, and hub nodes into one of three different roles.}. We do not consider within-community degree as our goal is not one of classification in their sense. Rather, we feel such a classification would detract from our primary focus on the role of gender within our networks. Another difference between our approach and that of \cite{GuimeraAmaral2005} is that we use the Girvan-Newman algorithm, which generates a hierarchy of partitions, as opposed to simulated annealing which maximises modularity. This decision should be discussed in some detail, for there are caveats to this experiment which we should stress.

In addition to the Girvan-Newman algorithm, we also used Louvain community detection and SparseBM. To ensure comparability among methods, Girvan–Newman, Louvain, and SparseBM were all applied to the corresponding zero-degree-removed graph. While modularity maximisation algorithms such as Louvain are popular, there are several flaws worth noting, cf. \cite{FortunatoHric2016}, \cite{GhasemianHosseinmardiClauset2019}. One is that modularity maximisation tends to overfit significantly. Another is that, paradoxically, it can also lead to systematic underfitting. These can occur simultaneously so that areas of the network that are dominated by randomness can be found to contain communities, while other areas with a clear modular structure can have such community structure obscured. This naturally leads to several questions concerning the validity of any output. While SparseBM avoids modularity maximisation, it fits a stochastic block model, whose communities are defined through stochastic equivalence\footnote{Roughly, two nodes are stochastically equivalent if they have the same probability distribution of ties with other nodes.}. In narrative networks, where central protagonists and minor characters may belong to the same narrative episode despite having markedly different structural roles, this modelling assumption may be less appropriate. Regardless, the additional use of these two algorithms demonstrated the sensitivity of the participation coefficient to the choice of communities. The three methods produced materially different participation coefficient distributions and, in some cases, different directions of the apparent gender contrast. The key is therefore to choose communities that more accurately reflect a `ground truth' of the network. Such communities are notoriously difficult to obtain and it is a significant challenge to validate whether obtained communities are in any way reflective of said `ground truth'. We argue the communities obtained via Girvan-Newman are a reasonable approximation of such a `ground truth', as we now explain.

First, we favour Girvan-Newman for its conceptual clarity. However, as is well known, it is a challenge to decide which partition of the Girvan-Newman algorithm is most suitable. We use modularity as a first attempt to choose such a partition. In particular, we look for two regions on a plot of modularity against partition: where the plot levels off; and where the plot reaches a peak. Such regions offer points of interest, as opposed to objectively appropriate partitions. Once we have identified such points of interest, we investigate the communities to analyse the extent to which they reflect the narrative.

\begin{figure}[t]
    \centering
    \includegraphics[width=0.5\textwidth]{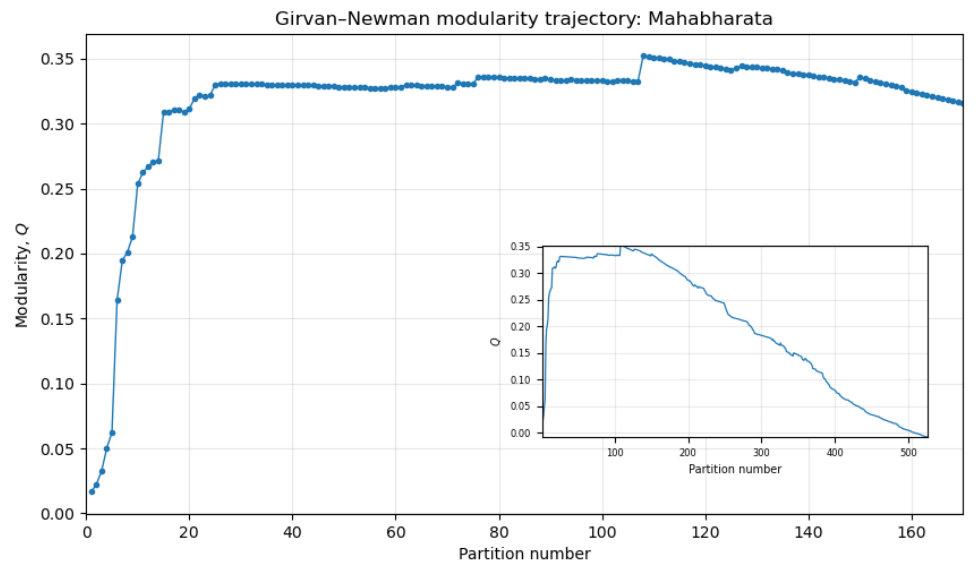}
    \caption{Modularity plotted against partition number for the \textit{Mah\={a}bh\={a}rata}. Inset is the full plot whereas the main plot offers a zoomed in perspective.}
    \label{fig:inset_partition_v_mod_maha}
\end{figure}

To identify a suitable partition in the \textit{Mah\={a}bh\={a}rata}, we first look to \figref{fig:inset_partition_v_mod_maha} which demonstrates two clear regions of interest. The first is at the start of the main plateau around partition 25, while the second is the peak at partition 108 in which $Q=0.3520$. We also investigate partitions 50 and 75 to add further detail to our investigation. The initial observation is that there is an explosion of singletons as the partition increases, with none at partition 25, 23 at partition 50, 41 at partition 75 and 63 at partition 108. There is also a very large, dominating community at partition 25 (with 222 members) that is slowly dismantled until it contains 124 members in partition 108. This suggests the start of the plateau is perhaps too coarse, and so we look for partitions along the plateau. In particular, when examining community members, we observe a community of 34 members which strongly reflects the \textit{R\={a}m\={a}ya\d{n}a} episode within the \textit{Mah\={a}bh\={a}rata}. This community remains intact through partitions 50 and 75, but is broken apart at some point by partition 108. We view this as a reasonably strong indicator of narrative suitability and therefore use it as a validation point. While still intact, we view successive partitions as broad refinements, while after it is broken apart we view further Girvan–Newman refinement as fragmenting otherwise coherent narrative groups, i.e over-refinement. We found that this validation criterion is met at partition 83. To summarise, we therefore have three partitions which may offer reasonably reflective community partitions: 25 (beginning of plateau), 83 (final partition before disintegration of \textit{R\={a}m\={a}ya\d{n}a} episode) and 108 (maximum modularity). We favour partition 83.

Before discussing the findings, we note that all three partitions produced reasonably similar results, and the qualitative conclusion is identical for all three partitions, i.e female characters consistently exhibit a higher mean participation coefficient than male characters (see below). Thus, the principal finding is not an artefact of a particular partition choice. We also noted that the absolute participation coefficients increase as the hierarchy is refined. This is expected, since successive Girvan–Newman partitions subdivide communities into progressively smaller groups, causing many nodes to distribute their neighbours across a larger number of communities and therefore increasing their participation coefficients. Most importantly, however, is that the increase in the gender difference occurs almost entirely between Partitions 25 and 83, with only marginal changes between Partitions 83 and 108. We argue this indicates that refining the hierarchy beyond the proposed narrative-preserving partition has little effect on the substantive comparison between male and female characters. The median participation coefficients reinforce this interpretation. We therefore view partition 83 as a particularly attractive compromise. It lies within the modularity plateau, preserves the integrity of the embedded \textit{R\={a}m\={a}ya\d{n}a} episode, and yields essentially the same substantive gender conclusions as the maximum-modularity partition while avoiding unnecessary fragmentation of the hierarchy.

\begin{figure}[t]
    \centering
    \includegraphics[width=0.5\textwidth]{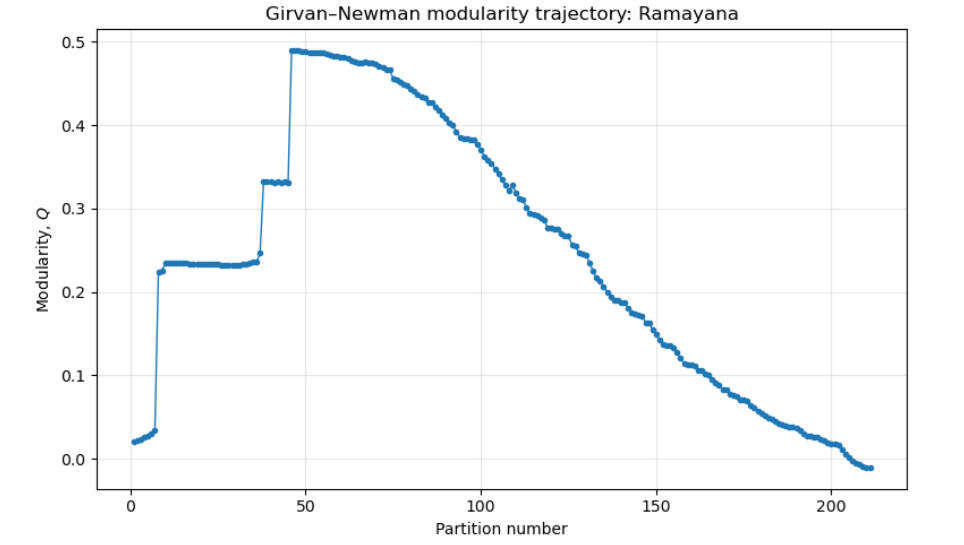}
    \caption{Modularity plotted against partition number for the \textit{R\={a}m\={a}ya\d{n}a}.}
    \label{fig:partition_v_mod_rama}
\end{figure}

For the \textit{R\={a}m\={a}ya\d{n}a}, the choice turned out to be clearer. As we see in \figref{fig:partition_v_mod_rama}, there are three regions of interest: first plateau (around partition 10), second plateau (around partition 38), and the maximum modularity (at partition 46). However, upon investigating community membership, a clear pattern emerged. Partition 10 seemed to be separating the Ayodhy\={a} court from the rest of the network. While appropriate, it was clearly far too coarse. Partition 38 then removed the Khara/D\={u}\d{s}a\d{n}a military substructures. Narratively this is also appropriate, though characters such as R\={a}ma and R\={a}va\d{n}a were still in the same community. As such, this is deemed too coarse still. Finally, partition 46 separates out the main protagonist from the main antagonist. Community 1 is now mostly R\={a}ma's alliance, while Community 3 is mostly made up of R\={a}va\d{n}a's court. This makes the most sense narratively and so we choose partition 46.

\begin{table*}[t]
\centering
\caption{Participation coefficient statistics by gender for nodes of degree at least four.}
\label{tab:participation_coefficient_geq4}

\begin{tabular}{l|l|cccccc}
\toprule
 & \textbf{Gender} &
Count &
Mean &
Median &
Std. Dev. &
Min &
Max \\
\midrule

\textit{Mah\={a}bh\={a}rata}
& Female
& 38
& 0.3314
& 0.3475
& 0.2880
& 0.0000
& 0.8000 \\

& Male
& 230
& 0.2088
& 0.0000
& 0.2695
& 0.0000
& 0.8760 \\

& Other
& 2
& 0.3600
& 0.3600
& 0.5091
& 0.0000
& 0.7200 \\

\midrule

\textit{R\={a}m\={a}ya\d{n}a}
& Female
& 18
& 0.2441
& 0.1994
& 0.1609
& 0.0000
& 0.6667 \\

& Male
& 114
& 0.2424
& 0.1726
& 0.2256
& 0.0000
& 0.8163 \\

\bottomrule
\end{tabular}
\end{table*}

With such a choice of communities, inspection of the participation coefficient rankings revealed several top ranked characters with low degrees. To remove undue influence from peripheral characters, we consider just those nodes whose degree is at least four. This threshold removes the most peripheral characters while retaining a sufficiently large sample for comparison. The results of this process are summarised in \tabref{tab:participation_coefficient_geq4}. Recall, the participation coefficient measures the extent to which a character's interactions are distributed across different communities. Characters with low participation coefficients interact predominantly within their own community, whereas those with high participation coefficients act as bridges linking multiple communities.

What we immediately see is that for both texts, there are characters attaining high participation scores. In particular, for both texts it is a male character who attains the highest participation score, with the \textit{R\={a}m\={a}ya\d{n}a} exhibiting the biggest difference between genders. For the \textit{R\={a}m\={a}ya\d{n}a}, we also see that the mean participation coefficients for men and women (likewise median) are approximately equal. This implies that while female nodes may be substantially under-represented among highly central nodes overall, those female characters who do act beyond a single community appear structurally comparable to their male counterparts in terms of cross-community distribution of ties. Male participation coefficients are more variable than female participation coefficients in this sample, with standard deviations of 0.2256 and 0.1609, respectively. This suggests that male characters occupy a broader range of community roles, although the relatively small number of female characters means that this comparison should be interpreted cautiously. The key takeaway is therefore a lack of gender asymmetry in general. Both genders typically distribute interactions across multiple communities rather than remaining confined to a single narrative module.

In contrast, the \textit{Mah\={a}bh\={a}rata} sees female nodes having substantially higher mean and median than male nodes. This suggests that among characters of degree at least four, female characters distribute their connections across communities more evenly than male characters. In other words, it appears as though women occupy disproportionately integrative positions within the narrative network. Note also the median of zero for male characters. This tells us that at least half of all male characters in the degree-at-least-four sample interact exclusively within a single community. Such characters are embedded within one narrative episode, or within strict narrative groups (e.g. military groupings) and do not connect otherwise separate parts of the narrative. If we contrast this with the female characters (whose median participation coefficient is 0.3475), this means a `typical' female character does not have all her interactions confined to one community. Instead, her interactions are distributed across multiple communities. A possible narrative explanation for this is a female character appearing across multiple narrative contexts through kinship, marriage, diplomacy, exile, abduction, motherhood etc. These create links between otherwise distinct communities. This offers a clear distinction from the \textit{R\={a}m\={a}ya\d{n}a}. As a final point on the \textit{Mah\={a}bh\={a}rata}, note that male and female characters have roughly similar standard deviations. As such, the observed gender difference rather reflects a shift in the central tendency of the distributions rather than greater heterogeneity among female characters.
\medskip

\noindent\textbf{Individual Structural Damage (Giant Component)}
\medskip

Next, we measure the reduction in giant-component size caused by deleting each individual node. We will then compare the distribution of these quantities for male and female nodes. We shall use the following measure:

\[
\Delta_i = |GC(G-i)|-|GC(G)|.
\]

\noindent We then compare $\Delta_i$ across genders. This allows us to examine node-level structural influence, rather than group-level attack. Instead of wondering what happens if we remove the top male/female nodes, now we ask for each individual node, what is the structural damage of its removal? So, whereas before we had to concern ourselves with how we are ranking nodes, what fraction/count we remove, now we can gain a much more granular insight into the importance of nodes categorised by gender. We may also gain insight that differs from standard centrality measures. For example, suppose a high degree node sits inside a clique. Locally, such a node is prominent. However, deleting it may not do much damage since many alternative routes remain. Note that we retain all characters in this analysis, including those outside the original giant component. This is intentional since exclusion of such characters would condition the comparison on giant component membership and thereby omit potentially meaningful gender differences in structural marginality. Consequently, a value of $\Delta_i=0$ may indicate that a character lies outside the giant component and therefore has no effect on its size when removed.

Consider \tabref{tab:gc_damage_summary}, in which we see the median change in giant component size is $-1$ for every category. This shows that the typical deletion reduces the giant component by one node. In other words, most characters are not structurally critical bridges.

For the \textit{Mah\={a}bh\={a}rata}, while the female and male means are somewhat close ($-1.1798$ and $-1.3341$, respectively), the standard deviations are substantially different, suggesting a broader spread in male structural importance. Moreover, male outliers exhibit much greater effects on giant component size. In other words, on average, deleting a male node produces somewhat greater structural disruption than deleting a female node, and the single most structurally important male node fragments the giant component substantially more than the most structurally important female node. Together, these results suggest that the \textit{Mah\={a}bh\={a}rata} contains a small population of highly influential male intermediary characters whose removal produces disproportionate structural damage. This remains consistent with our earlier robustness and betweenness analyses.

The \textit{R\={a}m\={a}ya\d{n}a} exhibits the same qualitative pattern. Male nodes again exhibit a larger mean decrease, larger standard deviation, and stronger extreme effects. Thus, the most structurally influential male nodes in the \textit{R\={a}m\={a}ya\d{n}a} again exert considerably greater influence on giant component connectivity than the most influential female nodes.

The results suggest that while female characters can certainly occupy locally important positions, in both networks the largest observed reductions in giant-component size follow the deletion of male characters, and mean damage is also greater among male than female characters. The larger male standard deviation further suggests that the male distributions contain a more pronounced upper tail of structurally consequential nodes.

\begin{table*}[t]
\centering
\caption{Summary statistics for changes in giant component size following single-node deletion.}
\label{tab:gc_damage_summary}

\begin{tabular}{l|l|cccccc}
\toprule
\textbf{Network} & \textbf{Gender} &
Count &
Mean &
Median &
Std. Dev. &
Min &
Max \\
\midrule

\textit{Mah\={a}bh\={a}rata}
& Female
& 89
& -1.1798
& -1.0
& 0.5947
& -5
& 0 \\

& Male
& 443
& -1.3341
& -1.0
& 1.3357
& -17
& 0 \\

& Other
& 4
& -1.0000
& -1.0
& 0.0000
& -1
& -1 \\

\midrule

\textit{R\={a}m\={a}ya\d{n}a}
& Female
& 39
& -0.8718
& -1.0
& 0.6147
& -3
& 0 \\

& Male
& 208
& -1.0817
& -1.0
& 1.1413
& -10
& 0 \\

\bottomrule
\end{tabular}
\end{table*}
\medskip

\noindent\textbf{Individual Structural Damage (Path Length)}
\medskip

Now we look at the effects on average path length. Unlike the previous giant-component analysis, which measured fragmentation, this measures changes in navigability and routing efficiency within the remaining connected core of the network. The summary of this can be found in \tabref{tab:apl_damage_summary}. What we see immediately is that for both texts, both the mean and median are approximately zero for both male and female characters, indicating that for the majority of nodes, deletion has very little effect on the global routing structure of the surviving giant component. This suggests that brokerage in this sense is concentrated in a relatively small set of nodes. With this in mind, if we consider standard deviation then we can see that in both texts there is wider distribution for male nodes than for female nodes. Moreover, for the \textit{Mah\={a}bh\={a}rata}, while the minimum values are broadly similar for male and female nodes, the maximum values are substantially larger for male nodes. This once again implies several extreme male nodes have greater influence on the network.

For the \textit{R\={a}m\={a}ya\d{n}a}, we again see this pattern. Thus, although the \textit{R\={a}m\={a}ya\d{n}a} is structurally more cohesive overall, the most influential routing nodes again appear disproportionately male. However, in this case there is an additional insight since we also have at least one male node who produces a much larger negative value. Recall, negative values indicate that deleting the node actually decreases the average path length of the surviving giant component. This may initially appear counter-intuitive, but need not be. For if we delete a peripheral or weakly connected node then this will remove distant low-connectivity regions from the giant component, thereby making the remaining core more compact and internally efficient. The much larger negative male minima may suggest that at least one male node sits on the boundaries between dense cores and weak peripheral regions. Removing such a node may therefore disconnect difficult-to-reach portions of the network, shortening average paths within the surviving core.

\begin{table*}[t]
\centering
\caption{Summary statistics for changes in average path length of the giant component following single-node deletion.}
\label{tab:apl_damage_summary}

\begin{tabular}{l|l|cccccc}
\toprule
\textbf{Network} & \textbf{Gender} &
Count &
Mean &
Median &
Std. Dev. &
Min &
Max \\
\midrule

\textit{Mah\={a}bh\={a}rata}
& Female
& 89
& -0.0015
& -0.0006
& 0.0075
& -0.0631
& 0.0077 \\

& Male
& 443
& 0.0009
& 0.0007
& 0.0140
& -0.0572
& 0.2145 \\

& Other
& 4
& -0.0018
& -0.0008
& 0.0057
& -0.0096
& 0.0040 \\

\midrule

\textit{R\={a}m\={a}ya\d{n}a}
& Female
& 39
& -0.0013
& -0.0002
& 0.0055
& -0.0190
& 0.0137 \\

& Male
& 208
& -0.0005
& 0.0000
& 0.0173
& -0.1011
& 0.1805 \\

\bottomrule
\end{tabular}
\end{table*}

Overall, we are building a consistent structural picture in which most nodes in both epics exert only local influence. However, there is a relatively small number of predominantly male characters which occupy disproportionately important structurally intermediary positions. This is especially prominent in the \textit{Mah\={a}bh\={a}rata}, where structural dependence on intermediary figures appears strongest. At the same time, the near-zero means and medians demonstrate that large-scale routing influence is highly concentrated rather than broadly distributed across the networks.
\medskip

\noindent\textbf{Structural Similarity}
\medskip

Finally, we compare female and male characters with broadly similar baseline structural properties. The preceding analyses either examined aggregate attacks on groups of highly central (male and female) nodes, or compared the distribution of single-node deletion effects across all characters. While informative, the former depended upon the choice of ranking measure and the number or proportion of nodes removed, while the latter compared gender-level summaries without accounting for differences in the structural positions occupied by individual characters. We therefore conduct a matched comparison designed to compare structurally similar male and female nodes.
\medskip

\noindent\textbf{Question:} For each female character, how do the single-node deletion effects compare with those of the structurally closest male character?
\medskip

To answer this, we first construct weighted distance networks. For each edge $(u,\,v)$, we define its total interaction strength by $w(u,\,v)=w_{\mathrm{pos}}(u,\,v)+w_{\mathrm{neg}}(u,\,v)$, and assign the corresponding effective distance $d(u,\,v) = 1/w(u,\,v).$ Thus, stronger interaction ties correspond to shorter effective distances. This reflects real-world networks in which those nodes with stronger connections more freely share information. Positive and negative interactions contribute equally to interaction strength in this construction.

Next, we compute the deletion effects for every node. Each female node is compared with its nearest eligible male match. For each node $i$, we compute:
\begin{itemize}
\item the change in giant-component size;
\item the change in weighted average shortest path length across all surviving connected components.
\end{itemize}
More precisely, we define

\begin{eqnarray*}
\Delta_i^{GC}
=&
|GC(G-i)|-|GC(G)|\\
\Delta_i^{APL}
=&
\ell_{\mathrm{pw}}(G-i)-\ell_{\mathrm{pw}}(G),
\end{eqnarray*}

\noindent where $GC(H)$ denotes the giant component of a graph $H$ and $\ell_{\mathrm{pw}}(H)$ denotes the pair-weighted average shortest path length across its connected components. If $C_1,\ldots,C_r$ are the connected components of $H$ containing at least two nodes, then
\[
\ell_{\mathrm{pw}}(H)
=
\frac{
\displaystyle\sum_{a=1}^{r}
|C_a|(|C_a|-1)\ell(C_a)
}{
\displaystyle\sum_{a=1}^{r}
|C_a|(|C_a|-1)
},
\]
where $\ell(C_a)$ is the weighted average shortest path length of $C_a$. Singleton components do not contribute to this average. We also define $GC_{\mathrm{damage},\,i}=-\Delta_i^{GC}$, so that larger positive values correspond to greater reductions in giant-component size.

Next, for every node we compute five baseline structural covariates:
\begin{itemize}
\item degree;
\item weighted degree;
\item triangle count;
\item betweenness centrality;
\item the size of the connected component containing the node.
\end{itemize}
The weighted degree is the sum of all positive and negative interaction counts incident to the node. Triangle count denotes the number of distinct three-character cliques containing the node, rather than the number of multiplicity-weighted triangle instances. Betweenness centrality is calculated on the unweighted network. To compare male and female nodes occupying similar structural positions, we perform nearest-neighbour matching from female to male nodes. Matching is exact with respect to giant-component membership: a female node belonging to the giant component can be matched only to a male node belonging to the giant component, while a female node outside the giant component can be matched only to a male node outside it. Within each of these two groups, matching is based on the five structural covariates.

If $x_i=(x_{i1},\ldots,x_{i5})$ denotes the covariate vector for node $i$, we first standardise each coordinate:
\[
z_{ij}
=
\frac{x_{ij}-\mu_j}{\sigma_j},
\]
where $\mu_j$ and $\sigma_j$ denote the mean and standard deviation of the $j$th covariate across all male and female nodes. For each female node $f$, we then calculate the Euclidean distance to every eligible male node $m$:
\[
d(f,m)
=
\sqrt{
\sum_{j=1}^{5}
\left(z_{fj}-z_{mj}\right)^2
}.
\]
For each female node, we identified all male nodes attaining the minimum Euclidean distance in the standardised matching space, treating distances $d$ satisfying $\lvert d-d_{\min}\rvert \leq 10^{-12}\max\{1,d_{\min}\}$ as numerically equal. Where several males were equally nearest, each  was assigned weight $1/k$, where $k$ was the number of co-nearest  matches, and their weighted mean outcome was used. Thus, each female character contributed one matched comparison. Matching is performed with replacement, so the same male character may serve as the closest comparison for more than one female character.

After constructing the matched pairs, we compare the following deletion outcomes:
\begin{itemize}
\item giant-component damage;
\item change in pair-weighted average shortest path length across the network.
\end{itemize}

\begin{table*}[t]
\centering
\caption{Matched male/female comparison summary for the \textit{Mah\={a}bh\={a}rata}. Differences are calculated as the matched male value minus the female value.}
\label{tab:maha_matched_summary}

\begin{tabular}{l|cccccc|ccc}
\toprule
\textbf{Metric} &
Count &
Mean diff. &
Median diff. &
Std. Dev. &
Min &
Max &
M greater (\%) &
Equal (\%) &
F greater (\%) \\
\midrule

GC damage
& 89
& 0.0787
& 0.0000
& 0.7107
& -4.0000
& 3.0000
& 8.99
& 87.64
& 3.37 \\

APL change
& 89
& 0.0007
& 0.0000
& 0.0085
& -0.0110
& 0.0725
& 50.56
& 2.25
& 47.19 \\

GC damage (w/out BC)
& 89
& 0.1613
& 0.0000
& 1.0361
& -3.5000
& 6.0000
& 17.98
& 69.66
& 12.36 \\

APL change (w/out BC)
& 89
& 0.0004
& 0.0000
& 0.0083
& -0.0162
& 0.0593
& 50.56
& 1.12
& 48.31 \\

\bottomrule
\end{tabular}

\tabletext{
Note: Here, ``Mean diff.'' denotes the average value of $\text{male value}-\text{female value}$ across matched pairs. For GC damage, positive differences indicate greater damage following
deletion of the matched male character. For APL change, positive differences indicate a more positive signed path-length change following
male deletion. GC-damage equality is exact, whereas APL outcomes are classified as approximately equal when the absolute matched difference does not exceed $\varepsilon=10^{-6}$.
}

\end{table*}

The summary of this experiment on the \textit{Mah\={a}bh\={a}rata} is in \tabref{tab:maha_matched_summary}. Recall, positive values mean male-favouring, while negative values are female-favouring. We consider the effects on the giant component first. Note that the mean is only slightly positive and the median is exactly 0, suggesting no strong systematic matched male advantage. Furthermore, the matched male has the larger value in just 8.99\% of pairs, again consistent with no strong male advantage. Tying this to our earlier observations, this suggests that there are a relatively small number of positive male-favouring differences. The distribution in fact turns out to be highly concentrated at zero with 87.64\% (78/89) of matched pairs exhibiting identical giant-component damage. This further reinforces the viewpoint that it is not a broad systematic asymmetry producing the small positive mean, but a relatively small number of positive male-favouring differences.

The strong concentration of matched differences at zero is also consistent with the view that much of the aggregate gender disparity is associated with differences in baseline structural position. Among characters closely matched on the selected baseline structural covariates, female nodes in the \textit{Mah\={a}bh\={a}rata} are generally at least as structurally influential as comparable male nodes. This produces a far more nuanced view than the earlier group-level attacks which suggested that removing highly central male nodes caused substantially more global disruption than removing highly central female nodes. The matched comparison now suggests that this difference arises primarily because male characters are more likely to occupy the most globally dominant structural positions in the first place. In other words, the principal asymmetry appears to lie in the distribution of structural positions rather than in substantial residual differences between approximately matched male and female characters.

The average path length results reinforce this interpretation. Recall that positive differences indicate a more positive signed path-length change for the matched male character. With this in mind, the mean difference is extremely small at 0.0007 and the median is zero to four decimal places. Using a tolerance of $\varepsilon=10^{-6}$, the male character has the more positive path-length change in 50.56\% of matched pairs, the two outcomes are approximately equal in 2.25\%, and the female character has the more positive change in the remaining 47.19\%. The male- and female-favouring proportions are therefore broadly balanced, providing little evidence of a systematic directional gender difference in weighted path-length effects. Because exact equality may be too restrictive when comparing floating-point quantities, we repeated the classification using several tolerances. With $\varepsilon=10^{-4}$, the proportions of male-favouring, approximately equal, and female-favouring pairs are 46.07\%, 8.99\%, and 44.94\%, respectively. Increasing the tolerance to $\varepsilon=10^{-2}$ gives corresponding proportions of 2.25\%, 96.63\%, and 1.12\%. Thus, as the tolerance is increased, a progressively larger proportion of the matched differences is classified as practically negligible. The $10^{-2}$ result should be interpreted cautiously since it treats all differences of magnitude at most 0.01 as equivalent. Across the smaller tolerances, the central finding is that male- and female-favouring signed differences occur at similar rates, with no clear directional asymmetry.

Taken together, \tabref{tab:maha_matched_summary} suggests that while the most globally central characters are predominantly male in the \textit{Mah\={a}bh\={a}rata}, female characters occupying comparable network positions can exhibit brokerage influence that is broadly similar to, and only rarely smaller or larger than, those of their matched male counterparts.

To assess the quality of the matching, we examined both the overall matching distances and the covariate-specific standardised differences. The matches were generally extremely close: the mean and median Euclidean distances were 0.0328 and 0.0003, respectively, while the $95^{th}$ percentile was 0.1697 and the maximum was 0.6032. This suggests that the larger discrepancies are confined to a small upper tail. Mean absolute standardised differences were below 0.017 for every covariate, and all pairs agreed exactly in giant component membership and component size. Although matching with replacement produced substantial reuse, the procedure distributed the comparison weight across a broad male pool: 216 distinct males were represented, of whom 155 appeared in more than one co-nearest match set. In total, 65.2\% of the male comparison weight was assigned to these reused males, but no individual male received a total effective weight greater than 3.0; that is, the equivalent weight of three complete female-level comparisons, or 3.37\% of the total male comparison weight.

\begin{table*}[t]
\centering
\caption{Matched male/female comparison summary for the \textit{R\={a}m\={a}ya\d{n}a}. Differences are calculated as the matched male value minus the female value.}
\label{tab:rama_matched_summary}

\begin{tabular}{l|cccccc|ccc}
\toprule
\textbf{Metric} &
Count &
Mean diff. &
Median diff. &
Std. Dev. &
Min &
Max &
M greater (\%) &
Equal (\%) &
F greater (\%) \\
\midrule

GC damage
& 39
& 0.0513
& 0.0000
& 0.3939
& -1.0000
& 2.0000
& 5.13
& 92.31
& 2.56 \\

APL change
& 39
& -0.0004
& 0.0000
& 0.0082
& -0.0424
& 0.0113
& 43.59
& 25.64
& 30.77 \\

GC damage (w/out BC)
& 39
& -0.0769
& 0.0000
& 0.2700
& -1.0000
& 0.0000
& 0.00
& 92.31
& 7.69 \\

APL change (w/out BC)
& 39
& 0.0006
& 0.0000
& 0.0079
& -0.0322
& 0.0268
& 43.59
& 25.64
& 30.77 \\

\bottomrule
\end{tabular}

\tabletext{
Note: Here, ``Mean diff.'' denotes the average value of
$\text{male value}-\text{female value}$ across matched pairs. For GC damage, positive differences indicate greater damage following
deletion of the matched male character. For APL change, positive differences indicate a more positive signed path-length change following
male deletion. GC-damage equality is exact, whereas APL outcomes are classified as approximately equal when the absolute matched difference does not exceed $\varepsilon=10^{-6}$.
}

\end{table*}

The corresponding results for the \textit{R\={a}m\={a}ya\d{n}a} are in \tabref{tab:rama_matched_summary}. The pattern here is very similar to the \textit{Mah\={a}bh\={a}rata}. The giant component damage results again show extremely small average differences, with mean difference 0.0513 and median difference zero. Only 5.13\% of matched male nodes produce greater giant component damage than their female counterparts, implying the slightly positive average is being driven by a small number of positive male-favouring differences. Furthermore, the standard deviation is 0.3939 and 92.31\% (36/39) of matched pairs are exactly zero. The small positive mean arises from only two positive male-favouring differences and one negative female-favouring difference.  This is consistent with the broader picture developed throughout this paper in which the \textit{R\={a}m\={a}ya\d{n}a} exhibits somewhat greater structural cohesion and less extreme intermediary structural effects than the \textit{Mah\={a}bh\={a}rata}. Taken together, these results once again suggest that if we control for baseline structural position, female nodes in the \textit{R\={a}m\={a}ya\d{n}a} are generally at least as structurally influential as comparable male nodes.

The average path length results are broadly consistent with the above, though there are caveats. The mean difference is extremely small at $-0.0004$ and the median is zero. However, using a tolerance of $\varepsilon=10^{-6}$, male characters have a positive path-length change in 43.59\% of matched pairs, while the two outcomes are approximately equal in 25.64\%, and the female character has the more positive change in the remaining 30.77\%. Thus, there is a male-favouring difference, though the high proportion of approximate ties does suggest that there is no substantial directional gender asymmetry. Again, because exact equality may be too restrictive when comparing floating-point quantities, we repeated the classification using larger tolerances. With $\varepsilon=10^{-4}$, the proportions of male-favouring, approximately equal, and female-favouring pairs become 28.21\%, 41.03\%, and 30.77\%, respectively. In particular, this difference arises from certain male-favouring pairs becoming approximately equal, indicating such male-favouring pairs were only slightly male-favouring. Increasing the tolerance further to $\varepsilon=10^{-2}$ gives corresponding proportions of 2.56\%, 94.87\%, and 2.56\%. Thus, as the tolerance is increased, a progressively larger proportion of the matched differences is classified as practically negligible. The $10^{-2}$ result should again be interpreted cautiously. A minimum difference of $-0.0424$, compared with a positive maximum of 0.0113, nevertheless shows that the most pronounced observed pairwise difference is female-favouring in its signed direction. Overall, the results indicate that weighted path-length differences between matched male and female characters are generally small, with no clear and consistent directional asymmetry.

The matching diagnostics for the \textit{R\={a}m\={a}ya\d{n}a} were also generally favourable. The mean and median Euclidean matching distances were 0.0893 and 0, respectively. The mean was higher than that for the \textit{Mah\={a}bh\={a}rata}, while the standard deviation of 0.2937 substantially exceeded the mean. Together with the zero median, the 95th percentile of 0.2549, and the maximum distance of 1.7784, these values indicate a strongly right-skewed distribution in which most matches were close but one pairing was exceptionally poor. In particular, the S\={i}t\={a}/Nala pairing accounted for much of the elevated mean and dispersion. Its large distance was driven principally by discrepancies in triangle count and betweenness centrality. Nevertheless, matching quality was strong overall. Mean absolute standardised differences remained below 0.048 for every covariate, and all female characters and their co-nearest male matches agreed exactly in giant-component membership and component size. Again, matching with replacement produced substantial reuse, though the procedure distributed the comparison weight across a broad male pool: 97 distinct males were represented, of whom 75 appeared in more than one co-nearest match set. Overall, 66.7\% of the male comparison weight was assigned to these reused males, yet no individual male received a total effective weight greater than 2.33, which is the equivalent of approximately 6.0\% of the total male comparison weight. This suggests that the results were not disproportionately influenced by any single male comparator.

As a final sensitivity analysis, we address a potential concern regarding the inclusion of betweenness centrality among the matching covariates. The initial experiment compares female and male characters occupying similar structural positions, and therefore estimates the remaining gender difference conditional on those characteristics. Because betweenness centrality is itself a measure of brokerage and is likely to be closely related to the node-deletion outcomes, matching on it could be regarded as overmatching. We therefore repeated the analysis after omitting betweenness centrality from the approximate matching covariates, while retaining degree, interaction strength, triangle count and component size, together with exact matching on giant-component membership. The results are also reported in \tabref{tab:maha_matched_summary} and \tabref{tab:rama_matched_summary}.

For the \textit{Mah\={a}bh\={a}rata}, omitting betweenness widened the differences slightly, primarily in the matched GC-damage differences. With betweenness included, 87.64\% of the comparisons had equal GC damage; without it, this proportion fell to 69.66\%. The aggregate numbers of male-favouring and female-favouring comparisons both increased by eight, reaching 17.98\% and 12.36\%, respectively. Thus, the reduction in equal outcomes was accompanied by more differences in both directions, although with a modest male tilt. The mean difference increased from 0.0787 to 0.1613, while the standard deviation increased from 0.7107 to 1.0361. The larger mean was influenced partly by the upper tail, with the maximum difference increasing from 3 to 6, whereas the median remained zero. Omitting betweenness therefore makes the male-favouring GC-damage contrast more pronounced, though the number of equal outcomes and female-favouring cases does add some nuance to this claim. More broadly, the comparison suggests that betweenness accounts for part of the observed gender asymmetry in GC damage, consistent with the importance of characters’ structural positions within the network. By contrast, the APL-brokerage results changed little. The mean difference remained close to zero, and the proportions of male- and female-favouring comparisons were closely balanced. There was therefore no consistent residual gender asymmetry in APL brokerage.

For the \textit{R\={a}m\={a}ya\d{n}a}, GC-damage outcomes remained overwhelmingly equal. Without betweenness, 92.31\% of comparisons were equal, none favoured the matched male comparator, and only 7.69\% favoured the female character. The mean consequently changed from a small positive value to a small negative one, but this reversal was generated by only three of the 39 comparisons and should not be interpreted as evidence of a systematic female advantage. The APL-brokerage results were also highly stable. Although the small mean difference changed sign, the standard deviations were similar and the proportions of male-favouring, approximately equal and female-favouring cases were identical under the $\varepsilon=10^{-6}$ tolerance. Thus, omitting betweenness altered the magnitudes of some APL differences but not their overall directional distribution.

Overall, it appears that the pronounced male advantage observed in the aggregate attack experiments is associated largely with the greater representation of male characters in globally central structural positions. Among the approximately matched pairs produced here, GC damage is usually very similar and signed weighted-path-length differences are generally small. We therefore find little evidence of a broad residual gender-associated difference in deletion effects after matching on the selected structural covariates.

\section{Conclusion}
\label{sec:conclusion}

In this study, we have continued the network examination of the \textit{R\={a}m\={a}ya\d{n}a} and the \textit{Mah\={a}bh\={a}rata} seen in \cite{Evans2026b}. Although the two texts differ in their narrative organisation, we have seen that they exhibit a broadly similar gendered structural pattern. Both networks are numerically dominated by male characters, and male characters are more strongly represented among highly connected, high-betweenness, and otherwise structurally prominent nodes. The imbalance is therefore not confined to the number of characters present but also extends to the distribution of influential network positions.

Nevertheless, the form taken by this asymmetry differs between the two texts. The \textit{R\={a}m\={a}ya\d{n}a} contains a smaller proportion of edges involving female characters. However, of those edges involving at least one woman, that edge is substantially more likely to connect two female characters than in the \textit{Mah\={a}bh\={a}rata}. The participation coefficient results also distinguish the two networks. In the \textit{Mah\={a}bh\={a}rata}, female characters exhibit higher participation coefficients on average, suggesting that they more frequently connect across community boundaries. In contrast, the \textit{R\={a}m\={a}ya\d{n}a} sees typical male and female characters having similar participation coefficients. More generally, the gender disparity in local connectivity is especially pronounced in the \textit{R\={a}m\={a}ya\d{n}a}, whereas the \textit{Mah\={a}bh\={a}rata} exhibits stronger disparities in betweenness centrality and shortest path intermediation.

Our node-removal experiments reinforce this distinction. Under both proportional and equal-count targeted attacks, removing highly central male characters produces greater disruption than removing the corresponding female characters. Male-node removal results in greater edge loss, larger reductions in giant-component size, more extensive fragmentation, and larger increases in average path length. The equal-count experiment showed that these differences cannot be attributed solely to the larger number of male characters removed under a proportional attack. We saw that the effect is strongest in the \textit{Mah\={a}bh\={a}rata}, consistent with the greater dependence of that network on structurally intermediary characters that we saw in \cite{Evans2026b}.

However, at the individual level, large deletion effects are uncommon. Structural influence is instead concentrated in a relatively small upper tail of nodes, with the largest observed effects produced by male characters in both networks. This initially appears to reinforce the broader picture of male structural dominance. The matched comparison, however, provides an important qualification. Once female characters are compared with male characters occupying closely similar positions, giant-component damage is usually identical and signed path-length differences are generally small.

Taken together, these findings suggest that the principal gender asymmetry in both texts lies primarily in the unequal distribution of structural positions. Male characters are more likely to occupy the most structurally consequential locations in the networks. Among approximately matched male and female characters there is little evidence of a broad residual difference in deletion-based structural influence. The distinction is therefore not simply between influential men and uninfluential women. Rather, it concerns unequal access to the network positions from which large-scale influence becomes possible.

This conclusion also clarifies a clear distinction between the two epics. The \textit{Mah\={a}bh\={a}rata} displays a more pronounced dependence on intermediary and community-spanning figures, while the \textit{R\={a}m\={a}ya\d{n}a} exhibits greater local cohesion and a somewhat more concentrated separation of narrative groups. We now see that gendered network structure reflects these broader differences: female characters in the \textit{Mah\={a}bh\={a}rata} act across communities more on average, whereas male and female characters in the \textit{R\={a}m\={a}ya\d{n}a} occupy more similar typical community-participation roles. The results therefore support the broader view that gendered structural patterns are shaped not only by numerical representation, but also by the distinct narrative organisation of each text.

\section{Appendix: Pronouncing Sanskrit}
\label{app:pronouncing_sanskrit}
 
When written in Roman script, Sanskrit words require the use of accented characters. While not strictly necessary for the above, it may be beneficial to understand some basic rules for such accented characters, given the oral history of these texts. We closely follow the relevant sections of \cite{BrockingtonRama} and \cite{SmithMaha}, and will only give a brief overview of what we have found useful when reading the aforementioned texts.

For vowels, a macron (e.g. \={a}, \={i}, \={u}) indicates a long vowel, thus:
\begin{itemize}
    \item `a' is similar to `u' in English `sun', while `\={a}' is similar to the first `a' in `saga';
    \item `i' is similar to the `i' in `sit', while `\={i}' is like the `ee' in `seem';
    \item `u' is like the `oo' in `soot', while `\={u}' is like the `oo` in `soon';
    \item `e' is like the vowel of `say';
    \item `o' is like the vowel of `so';
    \item The diphthongs `ai' and `au' are like the vowels of `sight' and `sound', respectively;
    \item In Sanskrit, `\d{r}' is a vowel which in North India is pronounced as `ri', while in South India it is pronounced as `ru'.
\end{itemize}
Note that the majority of names end in a vowel. Typically, male names end in a short vowel (-a, -i or occasionally -u), while female names end in a long vowel (-\={a}, -\={i}, -\={u}). Importantly, the length of a vowel is not related to the stress in pronunciation. All syllables are given equal stress.

For consonants, we have the following:
\begin{itemize}
    \item Dots below the letters `t,\,d,\,n,\,s' indicate retroflexion. In India, these are pronounced with the tongue curled upwards. For `s', this makes the sound similar to `sh' in `ship';
    \item An acute accent `\'{s}' on `s' represents a second, slightly different (non-retroflex) `sh' sound.
\end{itemize}
While there are other accents, including dots above letters, these can be largely ignored by the general reader.

To put the above into practice, K\d{r}\d{s}\d{n}a would be pronounced `Krishna' if using North Indian pronunciation, or `Krushna' if using South Indian pronunciation.

For unaccented characters, we have the following:
\begin{itemize}
    \item `c' is pronounced like `ch' in `chip';
    \item `h' is used in combination with many other consonants to indicate aspiration, e.g. `th' is not pronounced as in `the' but instead is pronounced more similarly to `t' or as in `goatherd' or `uphill'.
\end{itemize}

Finally, the first and final syllables of a name can be modified in various ways to represent the son or daughter of a character. For example, a son of Da\'{s}aratha is D\={a}\'{s}arathi, the son of R\={a}va\d{n}a is R\={a}va\d{n}i, and so on. A female example would be the daughter of Janaka being J\={a}nak\={i}.

\section{Acknowledgements}

The author would like to warmly thank P\'{a}draig MacCarron who kindly provided the \textit{Mah\={a}bh\={a}rata} dataset. The author also remembers with gratitude the late Ralph Kenna, who first introduced him to this area of research.

OpenAI's ChatGPT was used during manuscript preparation to review consistency between the manuscript and the accompanying Jupyter notebook and to assist in refining Python code, including consolidating analysis outputs into summary tables. The named author designed the study, wrote the initial code, performed and verified all analyses, interpreted the results, and takes full responsibility for the manuscript.

\section{Funding}
This work received no specific grant from any funding agency in the public, commercial, or not-for-profit sectors.


\printbibliography


\end{document}